\documentclass{aa}
\usepackage[utf8]{inputenc}
\usepackage{natbib}
\usepackage[colorlinks]{hyperref}
\usepackage[dvipsnames]{xcolor}
\usepackage{ulem}
\usepackage{booktabs}
\usepackage{dirtytalk}
\usepackage{physics}
\usepackage{comment}
\usepackage{soul} 

\hypersetup{
    citecolor=[RGB]{44, 117, 255},  
    linkcolor=[RGB]{231, 62, 1},
    urlcolor=[rgb]{0,0,1}
}

\usepackage[]{color}

\definecolor{red}{RGB}{255, 30, 0}

\definecolor{blue}{RGB}{0, 165, 255}

\definecolor{green}{RGB}{106,180,153}

\definecolor{violet}{RGB}{212,115,212}

\def\Trex{T-REx }

\def\Msun{M_\odot}
\def\Rvir{R_{vir}}

\graphicspath{{./figures/}}

\makeatletter
\renewcommand*\aa@pageof{, page \thepage{} of \pageref*{LastPage}}
\makeatother

\title{Tracing gaseous filaments connected to galaxy clusters: the case study of Abell 2744}

\titlerunning{}
\author{
  S.~Gallo\inst{1}\thanks{E-mail:~\tt{stefano.gallo@universite-paris-saclay.fr}},
  N.~Aghanim\inst{1}, 
 C.~Gouin\inst{2,1},
  D.~Eckert\inst{3}
   M.~Douspis\inst{1}, 
  J.~Paste\inst{4},
  T.~Bonnaire\inst{5}
  }
  
\institute{
Université Paris-Saclay, CNRS, Institut d’Astrophysique Spatiale, 91405, Orsay, France
\label{IAS}
\and
Sorbonne Université, UMR7095, Institut d’Astrophysique de Paris, 98 bis Boulevard Arago, F-75014, Paris, France \label{IAP}
\and
Department of Astronomy, University of Geneva, Ch. d’Ecogia 16, 1290 Versoix, Switzerland \label{UnivGeneva}
\and
Université Claude Bernard Lyon 1, 69100, Villeurbanne, France
\and
Laboratoire de Physique de l’École normale supérieure, ENS, Université PSL, CNRS, Sorbonne Université, Université Paris Cité, F-75005 Paris, France \label{ENS}
}
\date{July 15, 2024}

\abstract{
Filaments connected to galaxy clusters are crucial environments to study the building up of cosmic structures as they funnel matter towards the clusters’ deep gravitational potentials. Identifying gas in filaments is a challenge, due to their lower density contrast which produces faint signals. The best chance to detect these signals is therefore in the outskirts of galaxy clusters. 
We revisit the X-ray observation of the cluster Abell 2744 using statistical estimators of anisotropic matter distribution to identify filamentary patterns around it. We report for the first time the blind detection of filaments connected to a galaxy cluster from X-ray emission using a filament-finder technique and a multipole decomposition technique. We compare this result with filaments extracted from the distribution of spectroscopic galaxies, through which we demonstrate the robustness and reliability of our techniques in tracing a filamentary structure of 3 to 5 filaments connected to Abell 2744. 
}


\authorrunning{Gallo et al.}

\begin{document}

\maketitle

\section{Introduction}\label{sec:intro}

In the current picture of cosmic structure formation, matter in the Universe evolves under the effect of gravity, collapsing the potential wells generated by primordial density fluctuations. On large scales, matter forms a complex network of walls, filaments, and nodes called the Cosmic Web \citep{Bond1996-cosmic_web}. Galaxy clusters form at the nodes of the cosmic web, at the intersection of filaments that act as ``cosmic highways'' and funnel matter into the deep gravitational potential wells of the clusters. For this reason, the outskirts of galaxy clusters and the filaments connected to them are especially interesting regions to study the physical processes of matter accretion, such as mergers, shocks, and turbulence in gas motions \citep[see e.g., ][]{Reiprich2013-cluster_outskirts_review, Walker2019-cluster_outskirts_review, Walker2022-cluster_outskirts_review}. Furthermore, understanding the cluster mass distribution out to large radii is particularly important for their use as cosmological probes.

Therefore, understanding the distribution and properties of matter falling in clusters through filaments has been the goal of many studies, using hydrodynamical simulations \citep[e.g.][]{Rost2021-filaments_cluster-outskirts, Tuominen2021, Gouin2021, Gouin2022, Gouin2023, Galarraga-Espinosa2021, Galarraga-Espinosa2023}, stacked observations \citep[e.g.][]{deGraaff2019-stacked_filaments_Planck, Tanimura2019-stacked_filaments_Planck, Tanimura2020-stacked_filaments_ROSAT, Tanimura2022-stacked_filaments_eROSITA}, between cluster pairs or around single clusters \citep[e.g.][]{Werner2008-filament_A222-A223, PLANCK2013-filaments_cluster_pairs, Eckert2015-A2744, Bulbul2016_filaments_A1750, Bonjean2018-filament_A399-A401, Veronica2024-A3391/95_filaments}.

Nonetheless, detecting these large-scale filaments remains a difficult task, especially for the gas component. This is due to the lower density and temperature of filaments compared to clusters, which makes their signal fainter both in X-rays and Sunyaev-Zel'dovich (SZ) effect, and thus requires high sensitivity and low, stable noise. 

A particular case in which extended emission from filaments was observed in X-rays are the outskirts of the galaxy cluster Abell 2744 (A2744). \cite{Eckert2015-A2744} identified extended structures connected to A2744 from the adaptively-smoothed X-ray surface brightness map, and confirmed the detection using a sample of spectroscopic galaxies, as well as weak lensing observations.

Abell 2744 \citep[$z=0.306$][]{Owers2011-A2744_galaxies} is a very massive cluster \citep[$M_{200c} \sim 2 \times 10^{15} \Msun$][]{Medezinski2016-A2744_structure} and exhibits a highly disturbed dynamical state, with many different (up to eight) massive interacting substructures \citep{Kempner&David2004-A2744_xray, Merten2011-A2744_core_lensing, Owers2011-A2744_galaxies, Jauzac2016-A2744_structure, Medezinski2016-A2744_structure}. It has been extensively observed in X-rays, optical, and radio wavelengths \citep{Eckert2015-A2744, Owers2011-A2744_galaxies, Merten2011-A2744_core_lensing, Ibaraki2014-A2744_xray_outskirts, Jauzac2015-A2744_structure, Jauzac2016-A2744_structure, Boschin2006-A2744_galaxies, Kempner&David2004-A2744_xray, Rajpurohit2021-A2744_non-thermal_emission, Harvey&Massey2024-A2744_JWST_WL, Eckert2016-A2744_shock, Govoni2001-radio_halos, Hattori2017-A2744_filaments_WHIM, Medezinski2016-A2744_structure, Braglia2007-A2744_filaments_galaxies}. In the X-rays in addition to the two main peaks in the centre-south and towards the northeast, up to four cores have been detected \citep{Jauzac2016-A2744_structure}; moreover, the presence of density and temperature discontinuities suggests the presence of shocks \citep{Eckert2016-A2744_shock, Hattori2017-A2744_filaments_WHIM}. At radio wavelengths, the cluster hosts a large radio halo, as well as different radio relics in its outskirts (some of which are also associated with emission from shocks) \citep{Govoni2001-radio_halos, Rajpurohit2021-A2744_non-thermal_emission}.
Such a massive and unrelaxed cluster is thus the perfect candidate for being connected to many cosmic filaments, as predicted from simulations \citep[e.g.][]{Darragh-Ford2019, Gouin2021}.

In this work, we consider two automatic methods to identify filamentary structures in cluster outskirts: a 2-dimensional multipole decomposition \citep{Buote&Tsai1995-beta_power_ratio, schneider1997, Gouin2017, Gouin2019}, and the filament-finding algorithm \Trex \citep{Bonnaire2020-T-Rex, Bonnaire2022-T-Rex_math}. We apply these methods to the outskirts of A2744, using the same data as \cite{Eckert2015-A2744} (namely X-ray observations by \textit{XMM-Newton} and a catalogue of spectroscopic galaxies from \citeauthor{Owers2011-A2744_galaxies}, \citeyear{Owers2011-A2744_galaxies}) to obtain for the first time the blind detection of cosmic filaments connected to a galaxy cluster. Following \cite{Eckert2015-A2744}, we consider $\Rvir = 2.1\ h_{70}^{-1}$ Mpc, where $h_{70} = H_{0}/(70\, \mathrm{km}\, \mathrm{s}^{-1}\,\mathrm{ Mpc}^{-1})$.

The paper is organised as follows: Sect. \ref{sec:data} presents the data, together with the selection and preprocessing choices; In Sect. \ref{sec:methods}, we explain the details of the two methods we use; Sect. \ref{sec:results} is devoted to the presentation of the obtained results, first with the multipole decomposition method and then with the \Trex filament finder. In Sect. \ref{sec:robustness}, we discuss the robustness of our results and the impact of our choices; finally, in Sect. \ref{sec:discussion}-\ref{sec:conclusion} we discuss the implications of our results and present our conclusions.

In the analysis, we assume a flat $\Lambda$CDM cosmology with parameters from \cite{PLANCK2020-cosmological_parameters}, such that $H_{0} = 67.4 \, \mathrm{km}\, \mathrm{s}^{-1}\,\mathrm{ Mpc}^{-1}$, $\Omega_m = 0.315$, $\Omega_\Lambda = 0.6847$, and $\Omega_b = 0.0493$.

\section{Data}\label{sec:data}

\subsection{X-ray data}\label{sec:data_x-rays}

The cluster A2744 was observed by \textit{XMM-Newton} X-ray Observatory for 110 ks in 2014 \cite[see][for a detailed description]{Eckert2015-A2744,Jauzac2016-A2744_structure}. 
We re-analysed those data using the X-COP analysis pipeline \citep{Ghirardini2019-X-COP}.
We extracted raw images in two energy bands, $[0.4-1.2]$ keV and $[2-7]$ keV, along with the respective background and exposure maps. Point sources were detected in each band using the XMMSAS task \texttt{ewavelet} and the results cross-matched between the two energy bands.
This resulted in a first list of high-reliability point sources. We added to this list all the point-like sources in the soft images. This allows us to take into account the impact of the different depths in the soft and hard-band images. The final source list was used as a conservative mask to remove any potential signal not associated with that of the cluster and cluster outskirts. In practice, the areas covered by the identified point sources were masked and the corresponding pixels were refilled using Poisson realizations of the neighbouring surface brightness using the \texttt{dmfilth} tool of the \texttt{pyproffit} package\footnote{\url{https://github.com/domeckert/pyproffit}, \url{https://pyproffit.readthedocs.io}} \citep{Eckert2020}. 
For the purpose of this study, we decided to focus only on the structural properties of the X-ray emission. Therefore, we created what we call a ``hit map'', a binary map obtained setting to 1 every pixel in the (point-source filtered) surface brightness map whose value is above 0, i.e., lit pixels. 
Disregarding all information about the signal amplitude, the extended structures are highlighted as spatially concentrated collections of lit pixels, boosting the low signal structures with respect to higher-signal ones.
In the rest of the paper, this hit map is the data product that will be used in our analysis, and it will be referred to as X-ray map or X-ray data.

%
\subsection{Spectroscopic galaxies}\label{sec:data_galaxies}

In addition to the X-ray data, we considered the spectroscopic galaxy catalogue compiled by \cite{Owers2011-A2744_galaxies}. It gathers observation of A2744 with the AAOmega multi-object spectrograph on the Anglo-Australian Telescope, and catalogues from the literature \citep{Boschin2006-A2744_galaxies, Braglia2009, Couch&Sharples1987, Couch1998}. This compilation includes the redshifts of 1250 galaxies within 15 arcmin ($\sim 4$ Mpc) from the cluster centre.

Our analysis focuses on the cluster outskirts, probing the structures also along the line of sight. We thus define a rather large area to include all the galaxies around the cluster by considering the redshift range $cz_\text{cluster}\pm 5600 \, \mathrm{km}/\mathrm{s}$ \citep[with $z_\text{cluster}=0.306$,][]{Boschin2006-A2744_galaxies}. 
For the detection of the filamentary pattern (see Sect. \ref{sec:results_trex}), it is important to ensure spatial uniformity of the completeness so that the excess (lack) of galaxies in a particular region due to selection effects is not mistaken for a real local over(under)density. The spectroscopic completeness (within $\sim 11$ arcmin from the brightest cluster galaxy) was computed by \cite{Owers2011-A2744_galaxies} for different magnitude cuts, and it is shown in their Fig. 9. Based on this result, we chose for our analysis the galaxies with magnitudes $r_F<20.5$, in order to have the most uniform completeness possible. While there are still some differences in the completeness across regions, in particular towards the west of the cluster, most of the field is complete (with an overall completeness above 90\%).
The combination of the redshift and magnitude selection provides us with a catalogue of 305 galaxies, which we use to identify the filamentary structures connected to A2744.

In order to study the 3-dimensional structure of the galaxy distribution, we have corrected for the effect of peculiar velocities of galaxies within the cluster on the observed redshifts, in particular the Finger of God (FoG) effect \cite{Jackson1972-finger_of_god}. To correct for this redshift distortion, we relied on the assumption that a galaxy cluster has a galaxy distribution that is symmetrical along the line of sight and on the plane of the sky \citep[following][]{Tegmark2004, Tempel2012, Hwang2016}. The method we used proceeds in two steps as in \cite{Aghanim2024-Shapley}. The first step is to identify the affected galaxies thanks to a Friends-of-Friends (FoF) algorithm, using a linking length along the line of sight (LoS) equal to five times the one in the plane of the sky. In our case, we found that this method identifies all the galaxies we selected previously as part of the same FoF group\footnote{We also tested the FoF algorithm on a sample of galaxies with a wider range of velocities around A2744 ($\abs{cz}\lesssim20000\, km/s$), and found that the galaxies in our selection were in the same group, while the other galaxies were put in different groups in front or behind.}. 
Then in the second step, the LoS distances of the galaxies are compressed according to the group's elongation to remove the FoG distortion. The compression factor is computed as the ratio between the rms of the galaxy positions (w.r.t. the cluster centre) long the LoS and perpendicular to it.

%
\section{Analysis of A2744 outskirts}\label{sec:methods}

In this Section, we describe the two techniques used to identify filamentary structure around A2744, namely the aperture multipole decomposition and the T-ReX filament finder.

%
\subsection{Aperture Multipole Decomposition}\label{sec:methods_multipoles}

\begin{figure*}
    \centering
    \includegraphics[width = \textwidth]{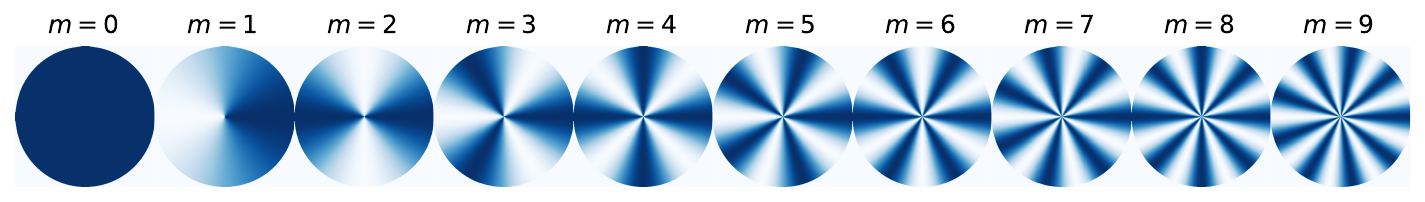}
    \caption{Illustration of the different angular symmetries associated to the different multipole orders $m$.}
    \label{fig:multipoles}
\end{figure*}

In the simplest approximation, galaxy clusters are considered spherical objects that accrete matter isotropically from their surroundings. We know from simulations and observations that this is generally not true, with clusters being in general triaxial objects \citep{Limousin2013}, and accretion happening preferentially via cosmic filaments connected to clusters \citep[e.g.][]{Reiprich2013-cluster_outskirts_review, Walker2019-cluster_outskirts_review, Walker2022-cluster_outskirts_review, Rost2021-filaments_cluster-outskirts, Gouin2021}. Still, we can expect that, in the outskirts of a galaxy cluster, the matter in the cosmic filaments falls into the gravitational potential well of the cluster in an approximately radial manner. With this approximation, which we expect to be valid within some radial range around the cluster, it is sufficient to study the azimuthal distribution of matter to identify filamentary structures connected to galaxy clusters.

To perform the analysis of the azimuthal distribution of gas and galaxies around A2744, we use the aperture multipole decomposition, a technique applied in various works to characterise the shape of galaxy clusters and their outskirts both in observations and simulations \citep{schneider1997, Dietrich2005, Mead2010, Shin2018, Gouin2017, Gouin2019, Gouin2022}.

The aperture multipole decomposition analyses the azimuthal behaviour of a 2D field in an aperture defined by the annulus $\Delta R = [R_\mathrm{min}, R_\mathrm{max}]$, where $R_\mathrm{min}$ and $R_\mathrm{max}$ are concentric radii. It consists of a decomposition of the field in harmonic orders $m$, each of them associated with a particular symmetry, as shown in Fig. \ref{fig:multipoles}. This technique is well suited for distinguishing the contribution of the different angular scales to the field considered, and by focusing on the lower (larger) multipoles one obtains information on the large-scale (small-scale) behaviour of the field. 

In practice for any 2D field $\Sigma (R,\phi)$ (where $\phi$ is the azimuth angle), the multipole of order $m$ in the aperture $\Delta R$ can be computed as:
\begin{equation}
     Q_m (\Delta R) = \int_{\Delta R} \ \Sigma (R,\phi) \ e^{im\phi} \ R \, dR  \ d \phi \, .
     \label{eq:multipole_qm}
\end{equation}

Given the integration in the radial aperture in Eq. \ref{eq:multipole_qm}, this decomposition is sensitive only to azimuthal patterns of the 2D field inside the aperture, and therefore is particularly adapted for fields that exhibit a radial symmetry, i.e. invariant along the radial direction. 

The multipole moments $Q_m$ being complex numbers, they contain information on both the power and the phase of the multipole of order $m$ in the field. Combining these two pieces of information for a number of orders, we can reconstruct the azimuthal structure of the original field up to a chosen order, that is, up to a chosen angular scale. Writing the multipoles as $Q_m = \abs{Q_m} e^{i \varphi_m}$, the reconstructed map up to order $m_\mathrm{max,rec}$ is given by

\begin{equation}\label{eq:reconstructed_map}
    \Sigma_\mathrm{reconst}\qty(\theta) = \sum_{m=0}^{m_\mathrm{max,rec}} \abs{Q_m}\, \cos\qty(m\, \theta - \varphi_m) \, .
\end{equation}

The choice of the maximum multipolar order to use in the reconstruction, $m_\mathrm{max,rec}$, is particularly relevant for the reconstructed map, as it selects the smallest angular scale included in the map. In this work, we are interested in the detection of large-scale cosmic filaments connected to A2744, and therefore we expect an extended signal spanning large angular scales, captured by low multipolar orders. On the other hand, we expect high multipolar orders (that is, small scales) to be dominated by signals from either point-like sources (in the case of X-rays) or small concentrations of galaxies. We also note that the multipole reconstruction assigns the same fraction of the azimuth to each component at every radius. For this reason, we chose $m_\mathrm{max,rec}=7$, which allows us to reconstruct large-scale structures, omitting the smallest-scale fluctuations \citep[see also][for details on the choice of multipole order limit]{Gouin2019}.

In order to compare the results of the multipole decomposition among different apertures or different fields, the values of $Q_m$ are not the most adapted quantities, as they depend on the normalization of the field, and it is the relative power of the order that gives information about the field structure. To get around this problem, we consider the multipolar ratios \citep{Buote&Tsai1995-beta_power_ratio, Gouin2022} between the modulus of the multipole of order $m$ and that of order zero:

\begin{equation}\label{eq:multipole_beta}
    \beta_m = \frac{\abs{Q_m}}{\abs{Q_0}}\,
\end{equation}

where $Q_0$ is the total amount of signal in the aperture and is given by $Q_0=\int_{\Delta R} \Sigma\, dA$. These ratios have two advantages over the use of $Q_m$: They are normalised making them comparable among different objects/apertures, and at the same time estimating the amount of power of the various multipoles with respect to the circular symmetry, thus probing the level of asymmetry of the considered distribution. For these reasons, they have been successfully used in the study of the morphology of galaxy clusters \citep[e.g.][]{Buote&Tsai1995-beta_power_ratio, Rasia2013-X-ray_cluster_morphology_estimators, Campitiello2022-xray_cluster_morphology, Gouin2022} and of galaxy cluster outskirts \citep{Gouin2022}.

When analysing the angular distribution of matter with aperture multipole moment decomposition, the two main parameters to consider are the position of the centre and the extent of the aperture $\Delta R$. It is therefore important to choose them so that the assumption of radial symmetry holds. 

In the perfect case of a spherical matter distribution, the natural choice for the centre would be the minimum of its gravitational potential. However, galaxy clusters are not perfectly spherical and the matter inside them is not homogeneously distributed. Indeed, anisotropic accretion processes, as matter flows from filaments, and merger events disturb their shape and matter distribution. Such processes create different offsets between the various matter components and the minimum of the potential. Centroid offsets are thus used as proxy to estimate the level of relaxedness of a cluster \cite[see e.g.][]{DeLuca2021}.

A2744 is a typical example of a very disturbed cluster, with many different substructures in the central area and at least two X-ray peaks  \citep[e.g.][]{Owers2011-A2744_galaxies, Jauzac2016-A2744_structure, Medezinski2016-A2744_structure, Harvey&Massey2024-A2744_JWST_WL}. It is therefore difficult to identify a clear and reliable centre for the azimuthal analysis.
Considering that we are interested in the matter distribution in the cluster's outskirts, we decided to circumvent this issue by examining the isocontours of the surface brightness map (smoothed with a Gaussian filter of size 7.5 arcsec). Indeed, we observed that lowering the signal threshold, the contours evolve from highly disturbed in the cluster centre to more regular, becoming roughly circular at about $1.6\times 10^{-6}\, \mathrm{counts}\; \mathrm{pixel}^{-1}\; s^{-1}$ (with a radius of about $0.6\, \Rvir \approx R_{500}$), before getting more irregular due to the filament emission. We therefore decided to centre the aperture $\Delta R$ in the middle of this contour, and chose $0.6\, \Rvir$ as its lower radius in order to exclude from the analysis the emission from the cluster itself, and focus just on the outskirts. 

The choice of the aperture's upper radius, $R_\mathrm{max}$, is, in general, dictated by the compromise between enclosing enough filamentary pattern signal and maximizing the signal-to-noise ratio within the aperture.
In the case of our data, the choice is also restricted by the extent of the data itself: for the X-ray data, $R_\mathrm{max}\leq 1.5\, \Rvir$; for the galaxies, $R_\mathrm{max}\leq 2.1\, \Rvir$. Therefore, for the X-ray case, we chose $R_\mathrm{max} = 1.4\, \Rvir$, to be symmetric around $\Rvir$. For the sparser galaxy data, instead, the need for more statistics motivated the choice of a larger aperture that includes all the available data, $R_\mathrm{max} = 2.1\, \Rvir$.

%
\subsection{\Trex filament finder}\label{sec:methods_t-rex}

An alternative way to characterise the distribution of matter around the clusters of galaxies is to determine the filaments connected to them. Among the numerous filament finder methods, we use for this study the Tree-based Ridge Extractor \citep[\Trex,][ to which we refer the readers for details]{Bonnaire2020-T-Rex, Bonnaire2022-T-Rex_math}. It is an algorithm specifically designed to identify one-dimensional filamentary structures in a discrete set of points embedded in higher dimensions (both in 2- and 3-dimensions). It is hence adapted for detecting cosmic filaments from a galaxy distribution.

The \Trex algorithm assumes a tree topology connecting the centroids of a Gaussian Mixture Model (GMM), which is then regularised to approximate the probability density function (pdf) from which the data points are drawn. In this way, the links between the centroids trace the ridges of the filaments in the data distribution. In practice, the first step of the algorithm is to build the minimum spanning tree\footnote{This is a graph with a tree topology connecting all the points in a dataset, so that the total length of its branches is minimised.} (MST) over the data. Then, the resulting tree is pruned by removing iteratively all end-point branches. This step is crucial to denoise the tree, removing all the small-scale branches at the extremities of the graph while retaining the relevant structures formed by long chains of connected centroids.
The output of this operation is hence a pruned tree that is used as prior for the regularised GMM (in which the nodes of the tree become the centroids of the Gaussians).
The GMM is then optimised over the full dataset so that it offers a smoothed representation of the tree that approximates the underlying distribution of the data.
To obtain a robust representation of the filament detection together with a measure of its uncertainty, the optimisation procedure described above is repeated $B$ times using a bootstrap approach. Each time, a subset of the data of size $N_B$ is chosen randomly, the MST is computed and then regularised. This produces a set of $B$ graphs, from which we can build a probability map of the filamentary structures, in which each pixel contains the number of times that a position is crossed by a tree branch. The role of each of the model parameters is discussed in \cite{Bonnaire2020-T-Rex}.

When applying \Trex to the 3D distribution of galaxies we use the following parameters: For the optimisation of the GMM we used values close to the ones suggested by \cite{Bonnaire2020-T-Rex}, in particular $l=2$ and $\lambda=1$; for the bootstrap method we perform $B=50$ iterations, each time sampling 90\% of the dataset, which corresponds to $N_B=275$.

Given that \Trex is optimised for discrete datasets, we needed to make some further adjustments to apply it to X-ray data. Therefore, we consider as input data points for the algorithm the centres of the lit pixels in the X-ray hit map, for a total of $N \sim 144\,000$ points. In the centre of the cluster (inside $\sim 0.6\, \Rvir$), though, basically all the pixels of the hit map are ``lit'', which means that we lose all information on the inner structure. Therefore, we decided to mask this area and focus only on the cluster outskirts (beyond $0.6\, \Rvir$).
In the remaining regions, the density of points is still very high, as is the noise. For this reason, we adapted the algorithm's free parameters to this different regime. For the GMM optimisation we used $l=200$ and $\lambda=100$. For the bootstrap, we chose $B=30$ and $N_B=0.7\, N\sim 101\,000$ points, to allow more variability between the various realisations and ensure convergence to the relevant structures.

%
\section{Results}\label{sec:results}

\subsection{Azimuthal distribution of matter}\label{sec:results_multipoles}

\begin{figure*}
    \centering
    \includegraphics[width = 0.9\textwidth]{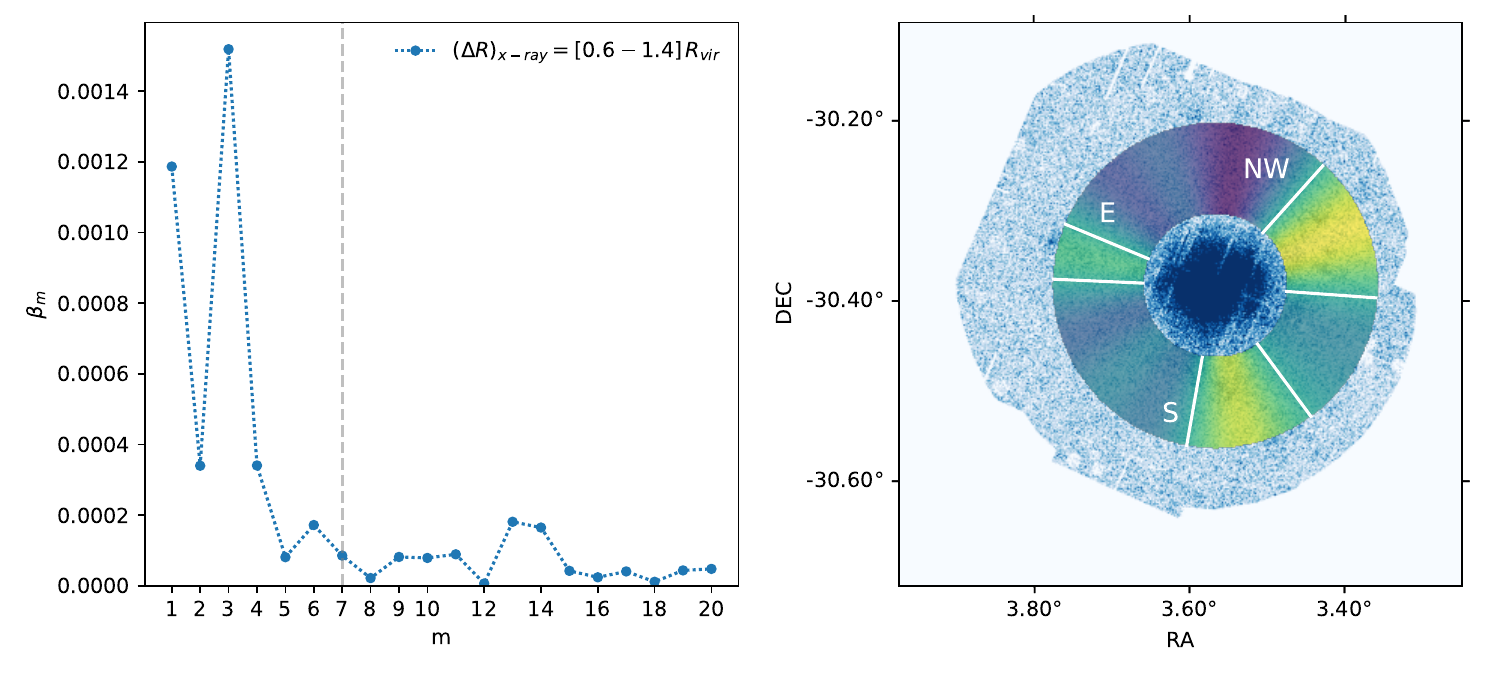}
    \caption{Left panel: Distribution of the multipolar ratio $\beta_m$ as a function of multipole order $m$, computed on one radial aperture $(\Delta R)_\mathrm{x-ray}=[0.6, 1.4]$. The limit order for the reconstruction, $m_\mathrm{max,rec}=7$, is shown as a vertical line. 
    Right panel: The reconstructed map in the aperture $(\Delta R)_\mathrm{x-ray}$. The white contours represent the threshold of 60\% of the map maximum, and identify the relevant filamentary structures, as described in the text. For reference, the X-ray hit-map is shown in the background.}
    \label{fig:Qm_06-14Rvir}
\end{figure*}

\begin{figure*}
    \centering
    \includegraphics[width = 0.9\textwidth]{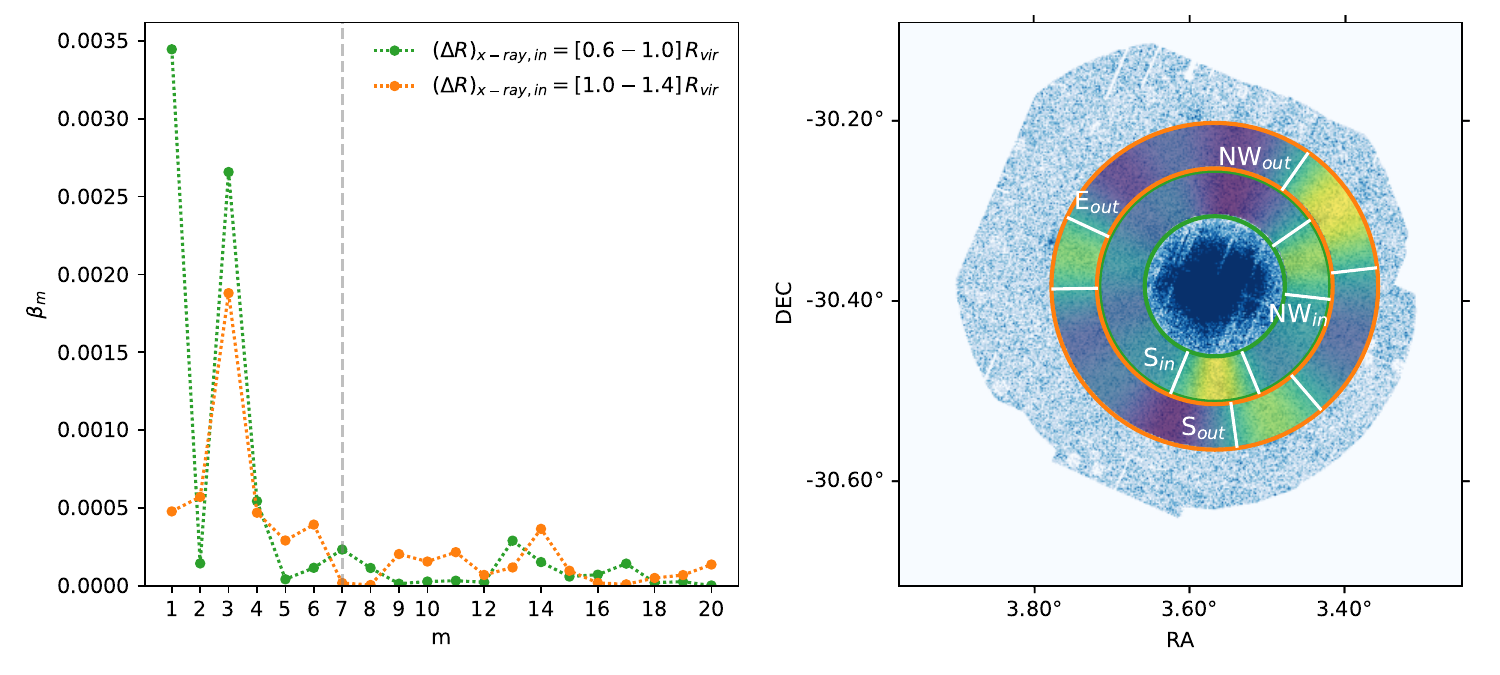}
    \caption{Same as Fig. \ref{fig:Qm_06-14Rvir} but considering two radial apertures, $(\Delta R)_\mathrm{in}=[0.6, 1.0]\, \Rvir$ (in green) and $(\Delta R)_\mathrm{out}=[1.0, 1.4]\, \Rvir$ (in orange).  }
    \label{fig:Qm_06-10-14Rvir}
\end{figure*}

\begin{figure*}
    \centering
    \includegraphics[width = 0.9\textwidth]{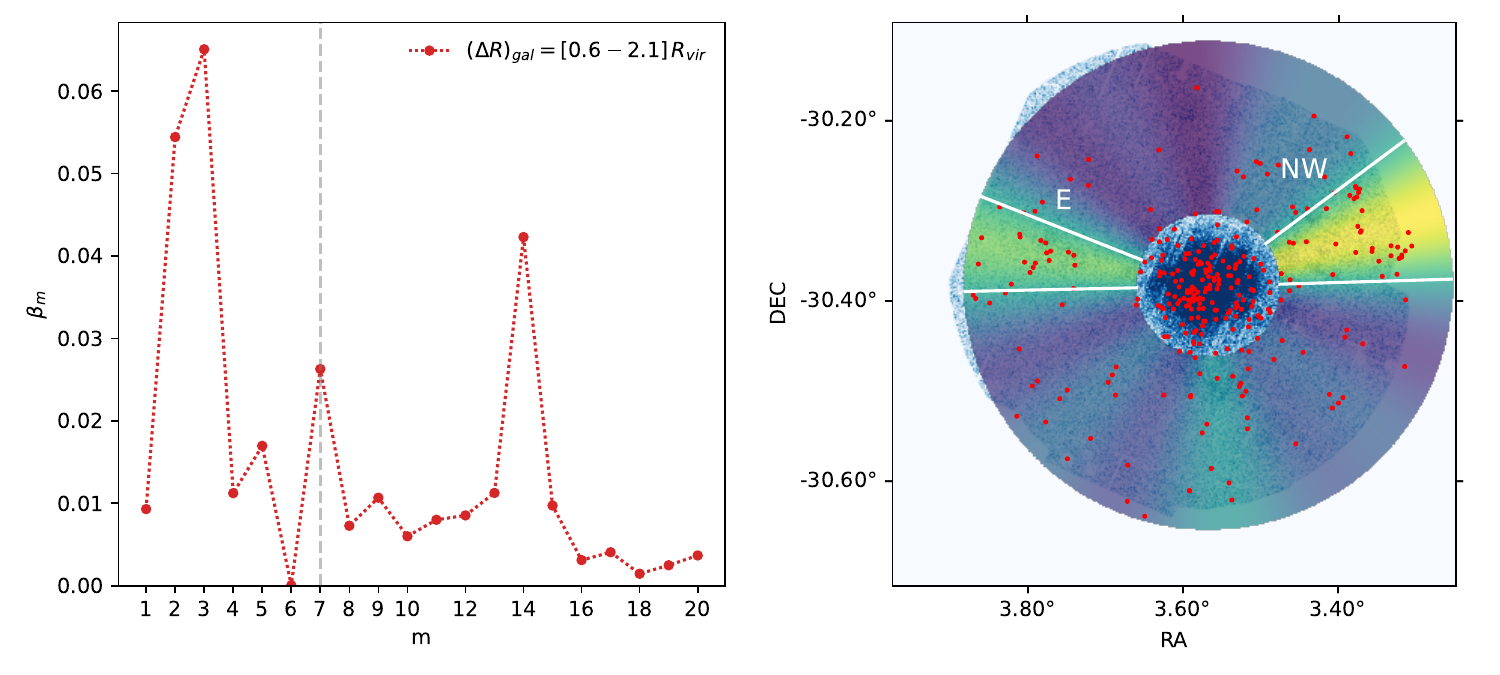}
    \caption{Same as Fig. \ref{fig:Qm_06-14Rvir} and \ref{fig:Qm_06-10-14Rvir}, but considering the 2D galaxy distribution inside one radial aperture $(\Delta R)_\mathrm{gal}=[0.6, 2.1]\, \Rvir$. }
    \label{fig:Qm_gal_06-21Rvir}
\end{figure*}

\subsubsection{Multipole decomposition of X-ray data}\label{sec:multipole_x-ray}

We first consider the azimuthal analysis of the X-ray data in a single large aperture $(\Delta R)_\mathrm{x-ray} = [0.6, 1.4]\, \Rvir$, in order to exhibit the large-scale azimuthal behaviour of the X-ray emission in the cluster outskirts. Then, we refine the analysis including multiple apertures, to account for a radial dependence of the structures.

In the left panel of Fig. \ref{fig:Qm_06-14Rvir}, we show the multipolar ratios $\beta_m$ for all orders up to $m=20$. 
One can see that the distribution of $\beta_m$ is not uniform, but clearly peaks at specific angular symmetries.
First, the octupolar symmetry (at the order $m=3$) dominates the distribution. Secondly, it is followed by the dipole ($m=1$), and then the even orders $m=2,4$. 
These angular symmetries represent structures at large angular scales, as expected for extended emission from cosmic filaments. 
Indeed, the symmetry at $m=3$ highlights the presence of three main structures (such as filaments) in the outskirts of A2744. 
In contrast, the dipolar signature shows an asymmetric signal between two halves of the aperture $(\Delta R)_\mathrm{x-ray}$ (as illustrated in Fig. \ref{fig:multipoles}). The combination of these two orders suggests that the X-ray data is mostly distributed into three structures, 
two of which are more prominent than the third (inducing this asymmetry).
Finally, the quadrupole order (at $m=2$) represents an elongated structure. It is often associated with the outer part of the halo elliptical shape \citep[e.g.][]{Gouin2017, Gouin2019, Gouin2022}.

In the right panel of Fig. \ref{fig:Qm_06-14Rvir}, we show the reconstructed map in the aperture $(\Delta R)_\mathrm{x-ray}$, obtained by summing the first seven orders of the multipole decomposition shown in the left panel.
In the map, we identify the relevant structures as the areas where the values are above 60\% of the map maximum. 
These regions are delimited by white contours in the figure.
We recognise three structures, which lie approximately in the northwest (NW), south (S) and east (E) directions. We see that, consistently with the expectations from the $\beta_m$ distribution, the NW and S structures are larger and stronger than the E structure. We also notice a mild alignment of the E and NW structures, due to the contribution of the even multipole orders ($m=2,4$).

By considering this large aperture $(\Delta R)_\mathrm{x-ray} = [0.6, 1.4]\, \Rvir$, we succeeded in capturing the broad distribution of structures in the outskirts of A2744. However, we lose the radial dependence information. To better recover the radial evolution, we analysed the data both inside and outside the virial radius, by splitting the aperture in two annuli: $(\Delta R)_\mathrm{in}=[0.6, 1.0]\, \Rvir$ and $(\Delta R)_\mathrm{out}=[1.0, 1.4]\, \Rvir$. We show the results of the multipole decomposition in the two sub-apertures in Fig. \ref{fig:Qm_06-10-14Rvir}.

Focusing on the inner aperture, $(\Delta R)_\mathrm{in}$, the $\beta_m$ distribution is dominated by the dipole, with the octupole as second-highest peak (Fig. \ref{fig:Qm_06-10-14Rvir}, left panel).
This inversion of the dominant order (compared to the single-aperture case) suggests a strong difference between two branches of the octupole and the third, which is confirmed by the reconstructed map (right panel). There, we see that only two structures are identified: one in the south (S$_\mathrm{in}$), and one in the west-northwest (NW$_\mathrm{in}$) direction. On the other hand, no structure is identified in the east of the aperture, where we would expect the third branch of the octupolar distribution. The reason why we cannot identify a structure in the east is that the X-ray signal in that area close to the cluster centre is weaker and extends over a larger angle compared to the other two regions. This may be related to a shock detected in the northeast region of the cluster \citep{Eckert2016-A2744_shock, Hattori2017-A2744_filaments_WHIM, Rajpurohit2021-A2744_non-thermal_emission}. 

Focusing on the outer aperture, $(\Delta R)_\mathrm{out}$, we find that $\beta_m$ is maximum at the order 3 (Fig. \ref{fig:Qm_06-10-14Rvir}, left panel). All the other orders up to $m=6$ are weakly contributing to the decomposition, with similar values of $\beta_m$. 
Examining the reconstructed map of the outer aperture (Fig. \ref{fig:Qm_06-10-14Rvir}, right panel), we clearly identify the three structures associated with the octupolar order, in the northwest (NW$_\mathrm{out}$), southwest (S$_\mathrm{out}$), and east (E$_\mathrm{out}$) directions. 
We note that the NW$_\mathrm{out}$ and S$_\mathrm{out}$ structures are both shifted counterclockwise with respect to their counterparts in the inner aperture, which highlights a radial dependence of these filaments across the cluster virial radius. We also see that, on top of the angular shift, NW$_\mathrm{out}$ has a larger angular size than NW$_\mathrm{in}$. On the other hand the sizes of S$_\mathrm{out}$ and S$_\mathrm{in}$ are similar.

\subsubsection{Multipole decomposition of galaxy distribution}\label{sec:multipole_galaxies}

Due to the low statistics of the galaxy sample (305 galaxies in total, including 150 in the inner cluster region within $0.6\, \Rvir$), we consider only one single large aperture, $(\Delta R)_\mathrm{gal} = [0.6, 2.1]\, \Rvir$. It contains all the 155 galaxies beyond $0.6\, \Rvir$ from the cluster centre. 
Fig. \ref{fig:Qm_gal_06-21Rvir} shows the $\beta_m$ of the galaxy distribution. The most important order of the decomposition is the octupole ($m=3$), identically to the X-ray data. 
We can notice that the quadrupole moment ($m=2$) is almost as strong as the octupole ($m=3$). This suggests that there are probably three structures, with two aligned on the same axis. 
Indeed, the reconstructed map (Fig. \ref{fig:Qm_gal_06-21Rvir}, right panel), shows that two main structures are identified in the east and west-northwest directions, almost opposite of each other, while the third one visible in the south is less significant, and does not cross our threshold of 60\% of the maximum. Nonetheless, we find this southern structure to be the third strongest of the reconstructed map.

Another feature is the high peak at $m=14$. This order traces very small angular scale structures. As discussed in \cite{Gouin2022}, large multipole orders are correlated with the fraction of substructures. Indeed, by including this order into the reconstructed map, we found that it is driven by small concentrations of galaxies (about $3-4$ galaxies with small angular separation). Notice that a similar excess of multipolar power at $m=13-14$ is also weakly significant for X-ray data (in Figs. \ref{fig:Qm_06-14Rvir}-\ref{fig:Qm_06-10-14Rvir}). This small-scale feature is not associated with large-scale filaments, but should rather trace sub-clumps of matter, and therefore is not considered further in our analysis.

%
\subsection{Filamentary structure around A2744}\label{sec:results_trex}

\begin{figure*}
    \centering
    
    \includegraphics[width=\textwidth]{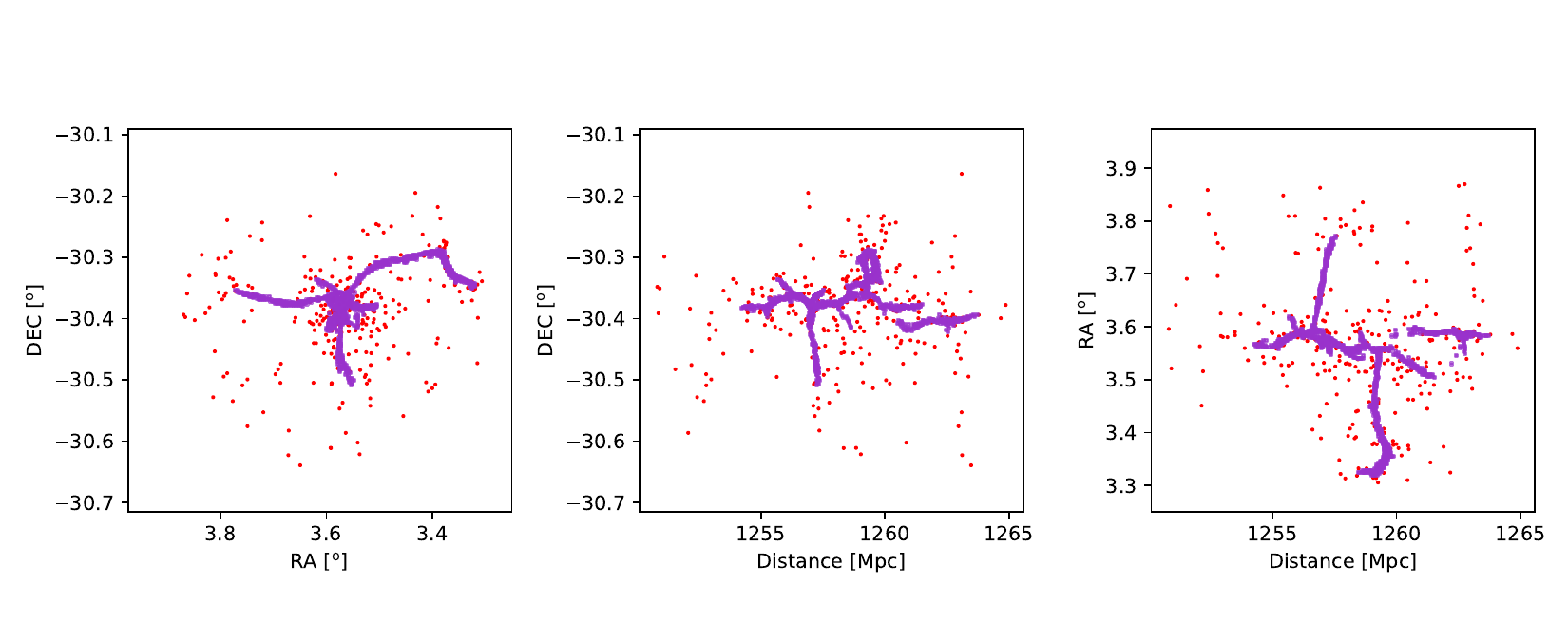}
    \caption{Three-dimensional distribution of galaxies (red points) in A2744, superimposed to the 3D probability map of the filamentary structures obtained with \Trex on the galaxy data. Only voxels with probability larger than 0.1 are shown. Left: projection along the line of sight. Middle and right: projections perpendicular to the line of sight, the viewer is on the left in both panels. 
    }
    \label{fig:trex_galaxies_probmap}
\end{figure*}

\begin{figure}
    \centering
    \includegraphics[width = 0.49\textwidth]{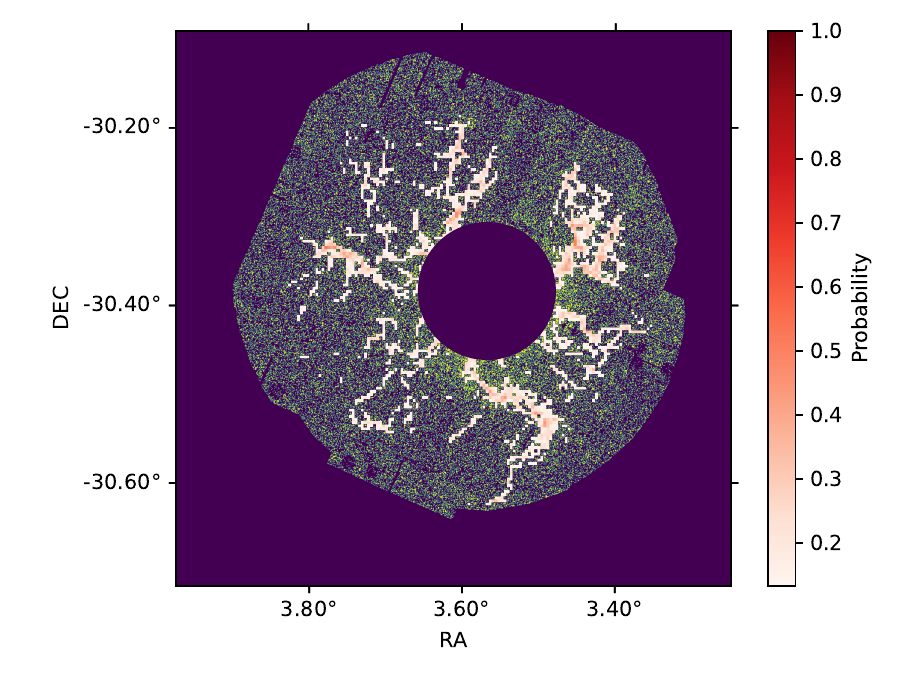}
    \caption{Probability map of the filamentary structures from X-ray data, obtained with \Trex. Only pixels with probability larger than 0.1 are shown. For reference, the X-ray hit-map is shown in the background.}
    \label{fig:trex_probmap}
\end{figure}

After analysing the angular distribution of X-ray signal and galaxy data through the multipole decomposition, we now focus on the filamentary structures around A2744, as detected by the filament-finding method \Trex. 

\subsubsection{Filaments from galaxy distribution}\label{sec:trex_galaxies}

With the spectroscopic galaxies, we can access the full three-dimensional (3D) information on the cluster and its environment. This allows the detection of structures along the line of sight, which results in a more accurate estimate of the cluster connectivity. We applied \Trex to the 3D distribution of galaxies, and we show the detected filamentary structures in Fig. \ref{fig:trex_galaxies_probmap}.

The panels in Fig. \ref{fig:trex_galaxies_probmap} depict three projections of the galaxy distribution, with superimposed the filament probability map thresholded at 0.1 (i.e. regions where the probability is higher than 0.1). The left panel shows the projection along the line of sight (with axes Right Ascension and Declination), we call this the face-on view. The central panel has the distance along the line of sight as $x$ axis and Declination as $y$ axis, which we call the side view. The right panel has again distance on the $x$ axis but Right Ascension on the $y$ axis, we refer to it as the top view. 

Starting from the face-on view, we can see that the \Trex algorithm identifies three filaments connected to the cluster (in the NW, E, and S directions), consistent with our previous results based on the X-ray data and with the results of \cite{Eckert2015-A2744}. We notice how the S filament is considerably shorter than the other two; this is in line with the results of the multipole analysis, where we found that in the south region the overdensity of galaxies is less significant than in the other zones. 
Another noticeable feature is the rather sharp bend in the NW filament at $\text{RA}\sim 3.5^\circ$, connecting a concentration of galaxies in the far west of the field. This is probably due to a selection effect, being so close to the edge of the observations.

Moving to the side view, we observe the cluster and its surrounding structures along the line of sight (on the $x$ axis). The first thing we notice is the presence of an elongated structure in the radial direction (horizontal in the picture), which extends both in front and behind the centre of the galaxy distribution, at relatively the same Declination. We call this the central structure.
Moreover, we observe that \Trex identifies a structure at the rear of the cluster, slightly separated from the main branch in the line of sight, which corresponds to the southern concentration of galaxies in the face-on view (at roughly $\text{RA}\sim 3.6^\circ$, Dec $\sim -30.4^\circ$). 
We can also see again the southern filament from a different point of view, and we see that it has very little extension in the line of sight direction, and that it connects to the central structure a little in front of the centre of the galaxy distribution.

Finally, from the top view we can extract some interesting information about the E and NW filaments. In fact, we see that the E filament connects to the central structure in approximately the same position as the S filament. On the other hand, the NW filament is connected further back, in a location that we can associate approximately with an X-ray peak in the northwest of the cluster centre. Moreover, we observe that both these filaments (actually all three considering also the S one) do not extend much in the radial direction, and tend to be roughly perpendicular to the line of sight. Lastly, we see that the central structure is not exactly parallel to the line of sight, but tends to extend from east to west as the distance increases, with the exception of the front-most part.

Putting everything together, we draw a picture of the three-dimensional structure of the A2744 cluster surroundings. We detect three filaments almost perpendicular to the line of sight; a long, extended filamentary structure along the line of sight, and a disconnected structure at the back of the cluster (slightly southeast of the centre). Two of the filaments (E and S) connect in the same position in the front-eastern part of the cluster, while the NW filament is connected in the back of the cluster, towards the west. 
Along the line of sight, the central structure extends beyond the virial radius both in front and behind the cluster: The front branch is located roughly at the centre of the cluster on the plane of the sky, almost parallel to the LoS; in the back of the cluster, the branch extends towards the west from the connecting point of the NW filament. 
Also in the back of the cluster, \Trex identifies another structure not directly connected to the main one, but crossing the virial sphere of the cluster. Located in the southwest of the cluster centre, this back structure can be associated with the southern peak in the X-ray surface brightness map.

\subsubsection{Filaments from X-ray data}\label{sec:trex_xrays}

We show the result of the \Trex algorithm in Fig. \ref{fig:trex_probmap}, where we see the \Trex probability map (only values above 0.1 are shown), superimposed to the X-ray ``hit map''. We identify large-scale, connected, and high-probability filaments. The most prominent of them is the filament in the S-SW. From the centre to the periphery, we see it connected to the cluster in the south of the masked area, then it extends to the south-west before exhibiting a bend to the south at RA$\sim3.5^\circ$; the last part of the structure shows reduced width and overall probability, so we do not trust it as reliable filament identification.
Another structure that we can clearly identify as a filament is the NW one: connected to the cluster in the W-NW, it extends to the NW before spreading and branching out in multiple directions between north and west. 
Moving the attention to the east of A2744, the identification of filaments becomes less straightforward. We can however distinguish two structures that stand out for their higher probability; both seem to depart from the same area in the E-NE of the cluster. One filament extends to the north and splits into two branches of lower reliability. The other filament extends to the east of the cluster. 
Finally, there are three regions where the probability is above the threshold of 0.1: west-southwest, southeast, and northeast. While we believe that these regions do not host real filaments, it might be interesting to investigate why some realisations of the \Trex algorithm identify some structures there. 
First of all, we should recall that this method was designed to work on much cleaner and sparser data, and that one of its main concepts is to connect overdense regions in a coherent tree structure. This means that if, in a particular bootstrap realisation, the optimisation identifies some local overdensity in the noise, it will tend to connect it with the overall tree. However, these spurious connections will not be stable across realisations, and will not tend to accumulate into a coherent structure in the probability map.
Looking at the surface brightness map, we can notice that, in all three regions, there seem to be small overdensities of signal close to the cluster outside the masked region. We think that these emissions drive the algorithm to connect noise structures along those directions, thus creating these broad, noisy probability distributions.
\par\bigskip
In summary, from the analysis of the \Trex results on X-ray data we identify four main structures as filaments extending to the south, northwest, east, and north directions. 

\begin{figure}
    \centering
    \includegraphics[width=0.49\textwidth]{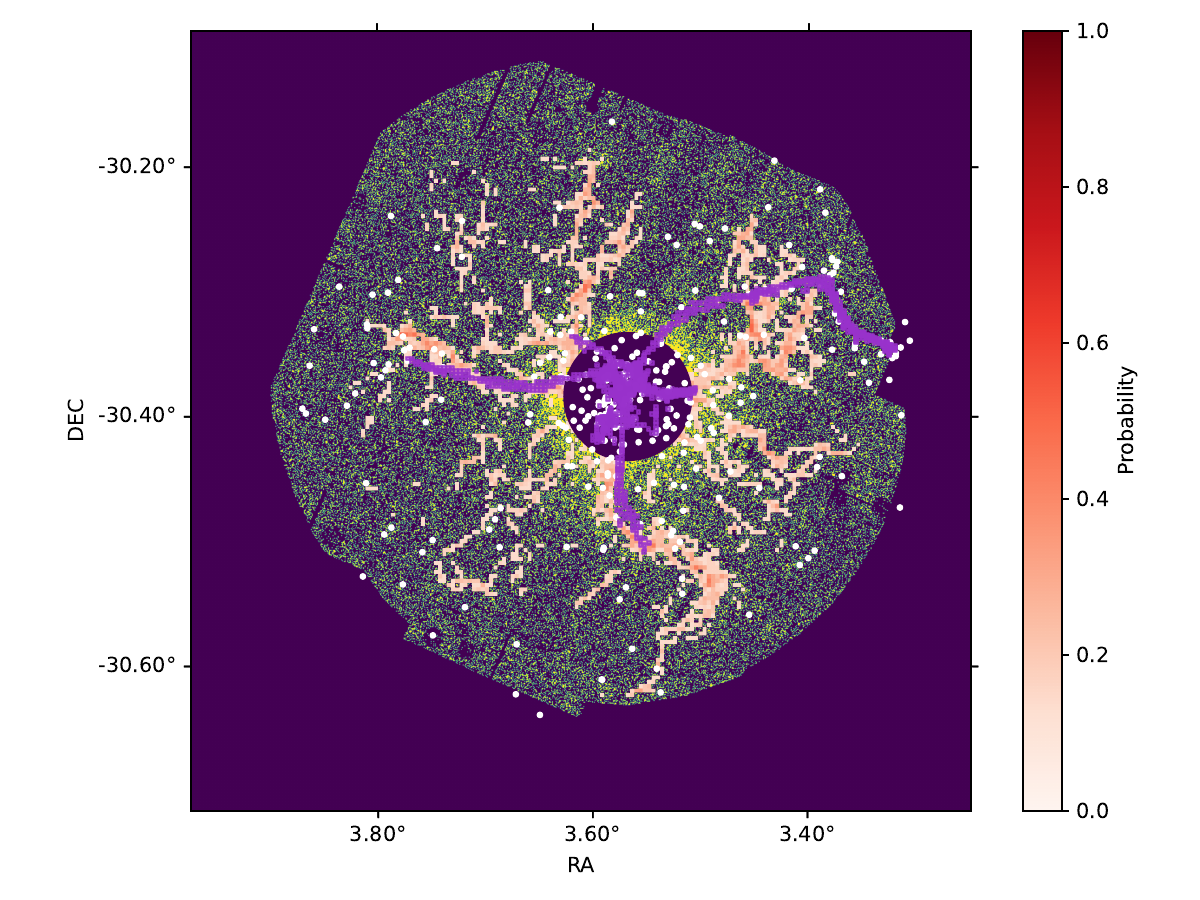}
    \caption{Comparison of the \Trex probability map from X-ray data (Fig. \ref{fig:trex_probmap}) and the face-on projection of the \Trex probability map from galaxy data (Fig. \ref{fig:trex_galaxies_probmap}, left panel). For reference, both the X-ray hit-map and the galaxy distribution are shown in the background.}
    \label{fig:trex_probmap_comparison}
\end{figure}

Comparing the probability map obtained from the X-ray data with that from the 3D galaxy distribution (Fig. \ref{fig:trex_probmap_comparison}), we note some important differences. The first one is the lack of a counterpart for the X-ray-detected N structure. This confirms the results of \cite{Eckert2015-A2744}, which also identified a northern structure in the X-ray map, but discarded it due to the lack of galaxies in the region (the X-ray emission was associated instead with a background galaxy concentration).
Another interesting difference is the length of the southern filament. In Fig. \ref{fig:trex_probmap}, we see it extends to a considerable length in the SW region, while in the galaxy case only the part close to the cluster is traced, out to $\sim 1\, \Rvir$. A possible explanation for this difference is a lack of galaxies in the southwestern region due to the lower completeness estimated by \cite{Owers2011-A2744_galaxies} (Fig. 9). The non-uniform spectroscopic completeness might also explain the different shape of the NW filament in the two \Trex probability maps, since the western region around $\Rvir$ is also reported to have lower completeness by \cite{Owers2011-A2744_galaxies}.

\section{Robustness of results}\label{sec:robustness}

To assess the robustness of the results presented in Sect. \ref{sec:results} we performed several tests, exploring different choices both in the data preprocessing and the parameters of the methods used.

\begin{figure*}
    \centering
    \includegraphics[width=0.6\textwidth]{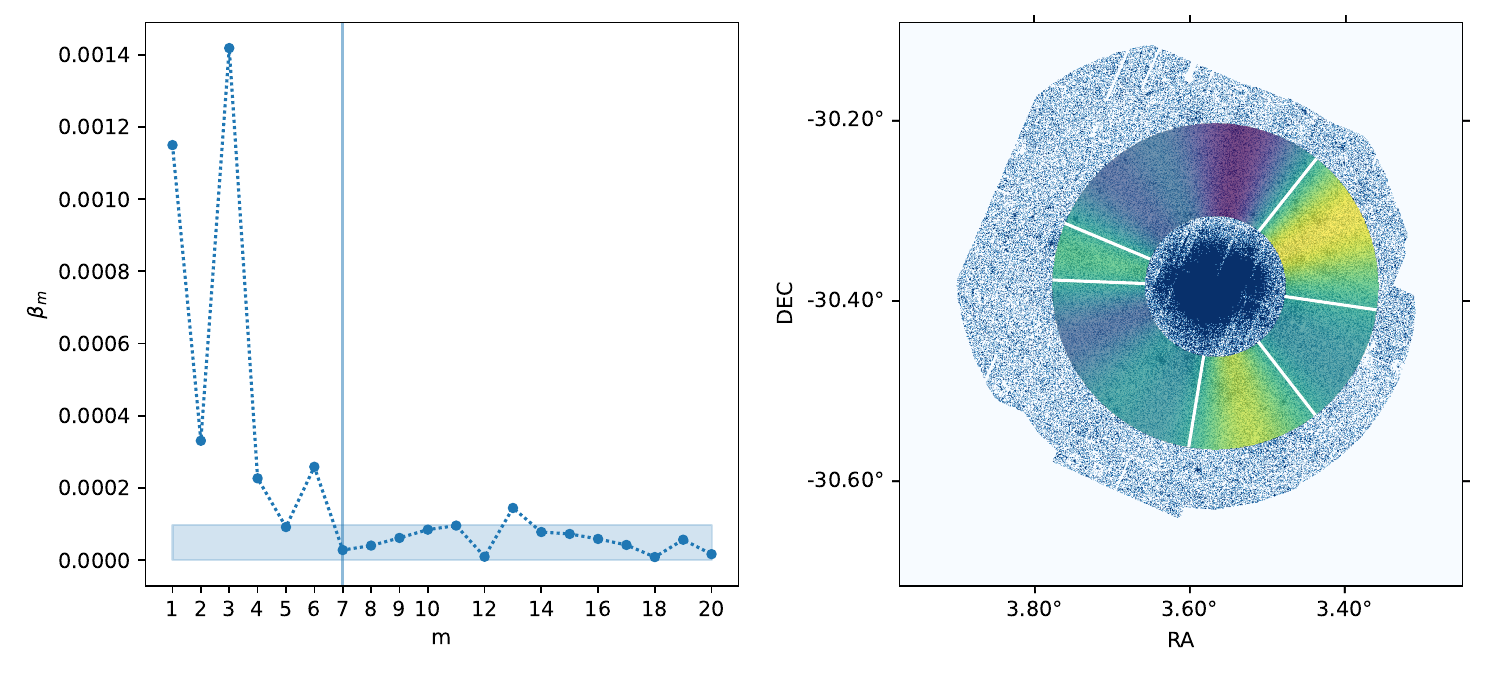}
    \includegraphics[width=0.36\textwidth]{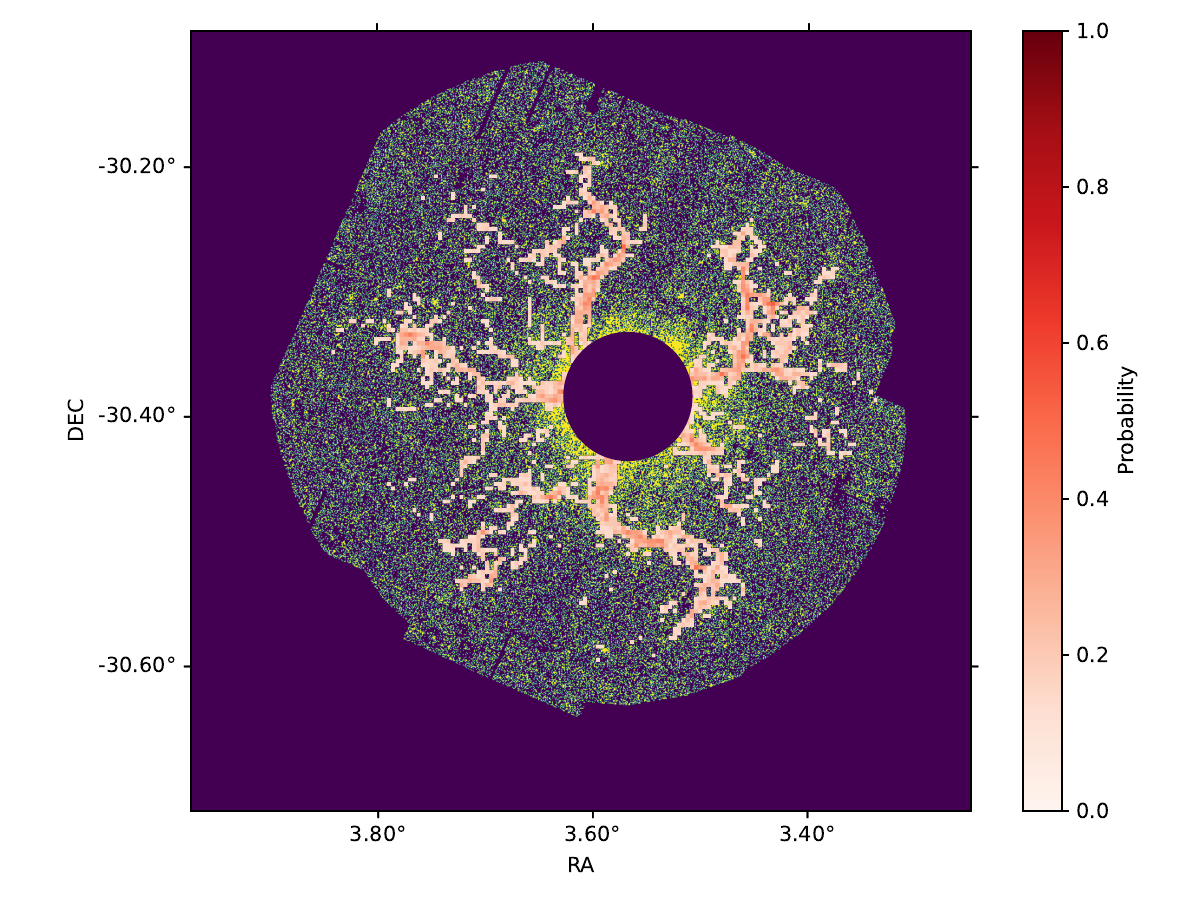}
    \caption{Results of the multipole decomposition and \Trex analyses, using the X-ray data with just the high-reliability point sources masked (see text). Left and middle: multipole analysis, same as Fig. \ref{fig:Qm_06-14Rvir}. Right: \Trex probability map, same as Fig. \ref{fig:trex_probmap}.}
    \label{fig:results_mmask}
\end{figure*}

\subsection{Robustness to data preprocessing choices}

We tested the impact of some of the choices described in Sect. \ref{sec:data}. For the baseline analysis of the X-ray data, we used a conservative point-source mask (see Sect. \ref{sec:data_x-rays} for details). We tested the impact of this choice by repeating both the multipole analysis and the T-REx filament detection masking only the high-reliability point sources; that is, those obtained by cross-matching the point sources lists detected in the soft and hard energy bands.
The results obtained from this set-up (Fig. \ref{fig:results_mmask}) are very similar to those of the baseline set-up (Figs. \ref{fig:Qm_06-14Rvir} and \ref{fig:trex_probmap}). We can notice some small differences in the $\beta_m$ distribution (Fig. \ref{fig:results_mmask}, left panel), but the reconstructed map (middle panel) does not change appreciably. The \Trex probability map (Fig. \ref{fig:results_mmask}, right panel) highlights the same four filaments as with the conservative point-source mask, and it looks somewhat less noisy, probably due to the alignment of some compact emissions with the filament signal, which helps the \Trex algorithm orient in those directions.

Moving on to the 3D galaxy distribution, one assumption in the baseline analysis is the magnitude cut at $r_F < 20.5$, based on the spectroscopic completeness. We explored the impact of this choice. In practice, we relaxed this constraint to $r_F < 21$, corresponding to 412 galaxies (35\% more than the baseline), and repeated the analysis. Again, the results are consistent with the baseline analysis. The reconstructed map from the multipole decomposition shows the same two main structures (E and NW) as the baseline analysis, while the southern one is also found below the threshold. The \Trex probability map is more complex than in the baseline analysis due to the larger number of galaxies, especially in the central region of A2744. 

%
\subsection{Robustness to method parameters}

We also tested the robustness of the analysis techniques, in particular concerning the choice of their free parameters. 
For the multipole analysis, we focused on the impact of the size of the aperture (i.e., the choice of the radial boundaries of the annulus). 
The choice of the aperture boundaries, $R_\mathrm{min}$ and $R_\mathrm{max}$, clearly influences the results, as shown for example in Fig. \ref{fig:Qm_06-10-14Rvir} when we compare two subsequent apertures. Nonetheless, we find that for the X-ray data in the single aperture case the qualitative results are quite stable across a large range of radial limits. For example, if we fix $R_\mathrm{max}$ at $1.4\,\Rvir$ and vary $R_\mathrm{min}$ from 0 to $1.2\,\Rvir$ we find that the most important order is always the octupole, while the relative power of the other orders vary, but the structures identified in the reconstructed map are always three, with small differences in position. On the other hand, fixing {$R_\mathrm{min}$ to $0.6\,\Rvir$ and varying $R_\mathrm{max}$, we find that between $0.8$ and $1.3\,\Rvir$ the dipole dominates over the octupole in the $\beta_m$ distribution, and in the reconstructed maps only the S and NW structures can be identified clearly (as for the inner aperture in Fig. \ref{fig:Qm_06-10-14Rvir}). After this radius, the octupolar order becomes dominant, and we find back the three filaments of the main analysis, mostly unchanged up to $R_\text{max}=1.5\,\Rvir$, which is the maximum distance we can reach before hitting the western border of the field.

The analysis of the galaxy data is significantly different in this respect. When dealing with a small number of galaxies, the multipole decomposition becomes particularly noisy, and the inclusion or exclusion of just a few galaxies significantly changes the results. Therefore, we need to use a large enough aperture to accumulate statistics. In the baseline analysis, we made the conservative choice of including all galaxies located beyond $0.6\,\Rvir$ from the cluster's centre. Fixing $R_\mathrm{min}=0.6\,\Rvir$ and progressively increasing the aperture size $\Delta R$, we first identify two structures in the south and northwest directions (up to $R_\mathrm{max}\sim1.5\,\Rvir$), then the southern one disappears, and later the eastern structure appears (around $1.7\,\Rvir$) and becomes more prominent as we expand the aperture to larger radii.

Regarding the \Trex algorithm, we conducted several tests to minimise the noise of the probability maps and ensure the best convergence of the trees. There are two main parameters we tested: the denoising parameter, $l$, and the strength of the tree prior, $\lambda$ \citep[described in detail in][]{Bonnaire2020-T-Rex}. The first parameter acts on the initial MST, on which an iterative pruning of all end-point branches is repeated $l$ times, denoising the tree while retaining the relevant long branches. The second parameter, $\lambda$, regulates the strength of the topological constraint in the optimisation, which tends to reduce the overall length of the graph. Higher values of $\lambda$ thus produce smoother and simpler filamentary structures (generally avoiding overfitting), but also tend to produce shorter branches, often at the expense of data fidelity.

\Trex has already been applied successfully on simulations to trace the cosmic web by \citep{Bonnaire2020-T-Rex, Gouin2021} and on 2D and 3D galaxy distributions in \cite{Aghanim2024-Shapley}. Therefore, in the present work, for the galaxy case we only slightly adapted the parameters, notably reducing the denoising parameter $l$ to account for the lower number of galaxies. 
For the case of the X-ray data, instead, we needed to adapt the parameters to the very different regime, with a large number of points and high noise level, as discussed in Sect. \ref{sec:results_trex}.  We found that a combination of high denoising $l$ and high $\lambda$ helps in reducing the noise in the probability map. In our tests, iteratively pruning the tree a large number of times proved useful in removing spurious detection. Looking at the onion decomposition  \citep[\citeauthor{Hebert-Dufresne2016} \citeyear{Hebert-Dufresne2016} as suggested by][]{Bonnaire2020-T-Rex} we found that choosing $l\sim200$ still allows for a large number of nodes to survive, thus allowing the tree to trace well peripheral areas. The choice of $\lambda$ is dictated by the need to reduce overfitting in the presence of a very large number of nodes, as explained in \cite{Bonnaire2022-T-Rex_math}.

Qualitatively, we find that the results of \Trex are consistent for a large range of the parameters' values, and that the regions we identified as filaments tend to have higher probabilities even in very noisy realisations of the probability map. This reduces the need to find the optimal free parameters.

%
\section{Discussion}\label{sec:discussion}

\begin{figure*}
    \centering
    \includegraphics[trim=300 0 450 0,clip,width = 0.49\textwidth]{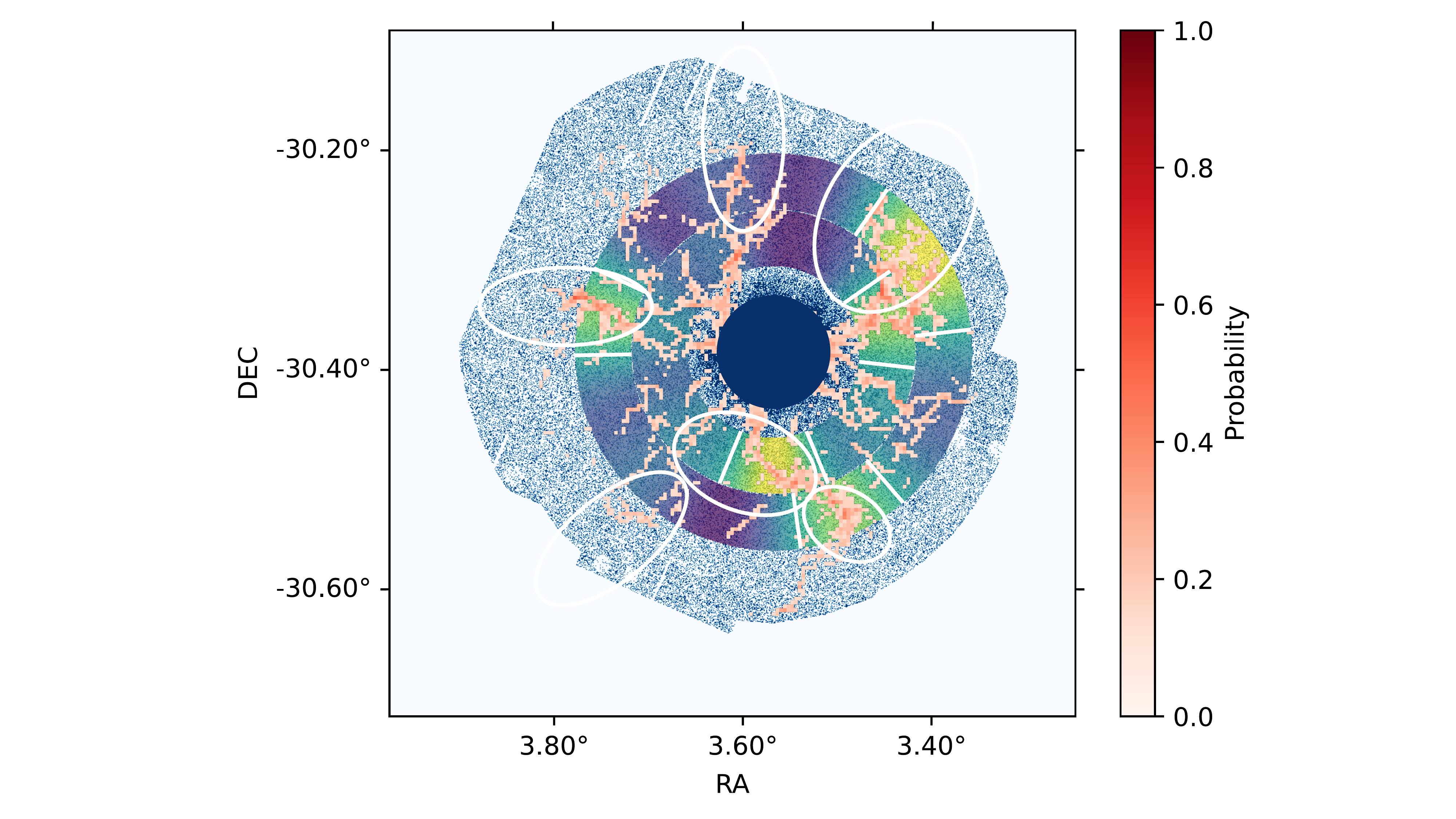}
    \includegraphics[trim=490 0 270 0,clip,width = 0.486\textwidth]{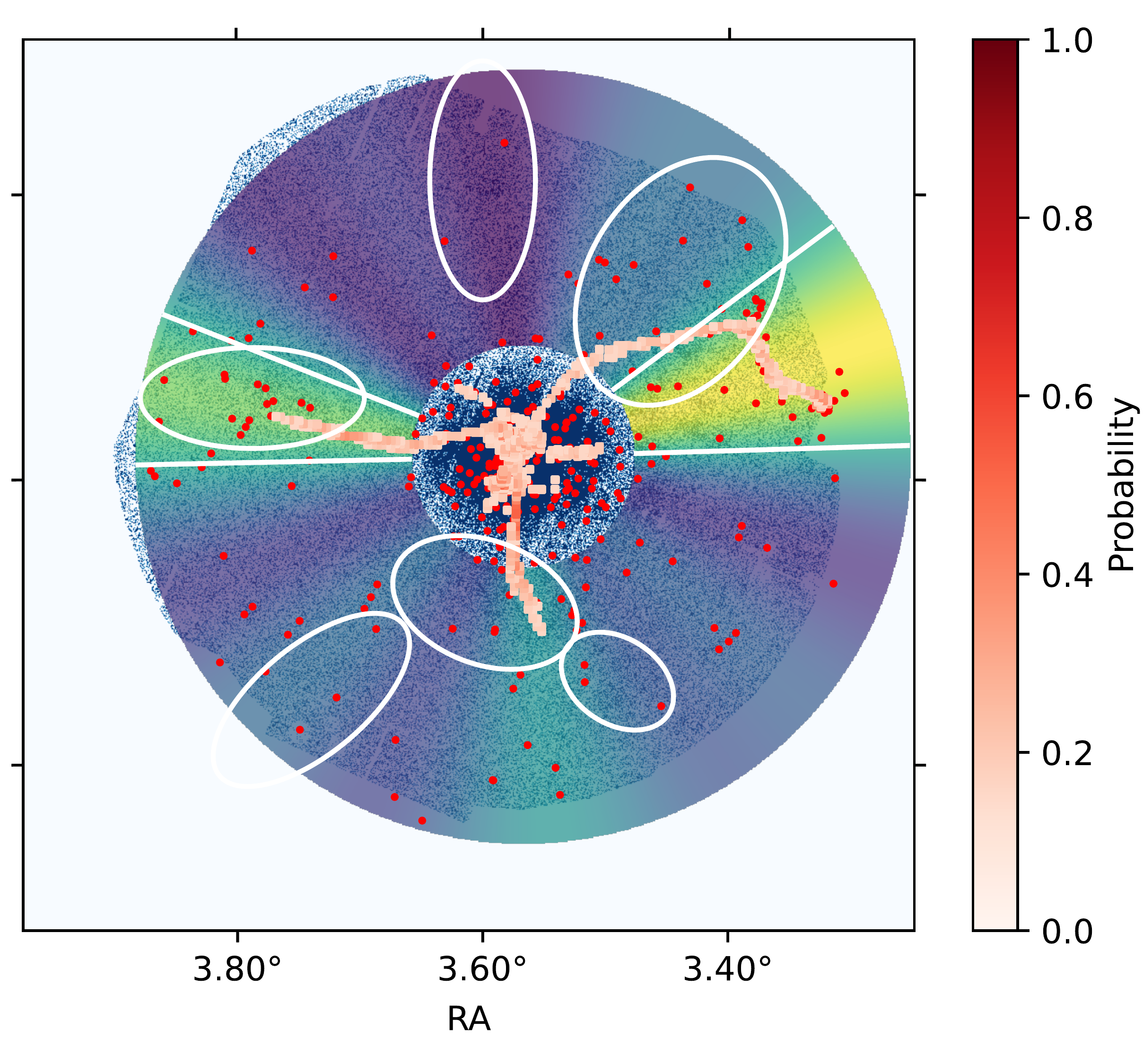}
    \caption{Results of the multipole and \Trex analyses, on X-ray data (left) and on galaxy data (right). The reconstructed map from the multipole decomposition is superimposed on the \Trex probability map. The white ellipses correspond to the regions identified in \cite{Eckert2015-A2744}. For reference, the X-ray hit-map is shown in the background.}
    \label{fig:comparison_Eckert}
\end{figure*}

The outskirts of the cluster of galaxies A2744 were the target of different studies, that probed them with spectroscopic galaxy and X-rays observations \citep{Braglia2007-A2744_filaments_galaxies, Owers2011-A2744_galaxies, Ibaraki2014-A2744_xray_outskirts, Eckert2015-A2744, Hattori2017-A2744_filaments_WHIM}.
\cite{Braglia2007-A2744_filaments_galaxies} used a sample of 194 spectroscopic galaxies observed with the VIsible MultiObject Spectrograph at ESO's Very Large Telescope (VLT-VIMOS), while \cite{Owers2011-A2744_galaxies} used the same sample of galaxies we used in this work. Both studies combined position and velocity information, finding two overdense regions in the south and northwest of A2744 which were identified as large-scale filaments. \cite{Owers2011-A2744_galaxies} also highlighted an overdensity of galaxies in the east of the cluster, but found that the local velocity distribution does not differ significantly from the overall cluster distribution, so it was not identified as a relevant structure by the authors. 

\cite{Eckert2015-A2744} identified six regions of extended emission connected to the cluster from the adaptively-smoothed X-ray surface brightness map (in the soft energy band). Four of these filamentary structures were found to coincide with galaxy concentrations: one in the northwest (NW$_{E15}$), two in the south and southwest (S+SW$_{E15}$), one in the east (E$_{E15}$). The first two regions correspond to the findings of \cite{Braglia2007-A2744_filaments_galaxies} and \cite{Owers2011-A2744_galaxies}, the east one to the overdensity of \cite{Owers2011-A2744_galaxies}. The other two X-ray emission regions (in the southeast, SE$_{E15}$, and north, N$_{E15}$) were associated with concentrations of galaxies not connected to A2744.
Analysing the X-ray spectra in the detected regions, \cite{Eckert2015-A2744} and \cite{Hattori2017-A2744_filaments_WHIM} found evidence that the gas originating the X-ray emission is in the form of Warm-Hot Intergalactic Medium (WHIM). This supports the identification of these regions as cosmic filaments, since several studies showed that, in hydrodynamical simulations, the WHIM is the most important gas phase in cosmic filaments, and can be reliably used to trace them \citep{Martizzi2019, Galarraga-Espinosa2021, Tuominen2021, Gouin2022}. Furthermore, again from simulations, \cite{Gouin2023} showed that the dominant source of soft X-rays beyond the virial radius is the warm gas in the WHIM and warm circumgalactic medium (WCGM), instead of the hot gas of the intra-cluster medium (ICM) that dominates inside $\Rvir$.

In this work, we performed the first blind detection of filaments around A2744, using a filament-finder technique and reconstructed maps from a multipole decomposition technique, applied both to the X-ray data and to the galaxy distribution.

In Fig. \ref{fig:comparison_Eckert}, we compare the structures identified with our two methods, for X-ray (left panel) and galaxy (right panel) data, with the X-ray-identified structures (white ellipses) from \cite{Eckert2015-A2744} (and thus, implicitly, with the other works mentioned above). From the X-ray data, we see that for the NW, S, and E structures there is a very close correspondence between the results of our two methods. The detected filaments from \Trex match well the region identified in the reconstructed maps, both in terms of position and radial dependence, including the gap between the cluster and the E filament. These structures are also well in agreement with the NW$_{E15}$, (S+SW)$_{E15}$ and E$_{E15}$ regions from \cite{Eckert2015-A2744}. The fact that these regions (the ones with confirmed galaxy counterparts in \cite{Eckert2015-A2744}) are the ones where our methods agree highlights the complementarity of the two methods, and the robustness of their combined detections. We also note that the N filament, detected by \Trex, has no significant match in the multipole analysis, but it does match with a structure (N$_{E15}$) first identified and then discarded by \cite{Eckert2015-A2744}. Finally, the SE$_{E15}$ ellipse has no detected counterpart in either of our two methods.

For the galaxy data, we recall that the \Trex algorithm is run on the 3D distribution of galaxies, while the multipole analysis is done on the projected 2D distribution. Comparing the results of the two methods along the line of sight (Fig. \ref{fig:comparison_Eckert}, right panel), we see that: in the east, the \Trex-detected filament overlaps well with the E structure in the reconstructed map and with the E$_{E15}$ ellipse. In the northwest, the overlap between the different detections is only partial, but they all agree on the presence of a structure in this area. The reasons for these differences might be selection effects (due to the non-spatially-uniform completeness of the galaxy sample) or border effects. In the south, as we mentioned in Sects. \ref{sec:multipole_galaxies}-\ref{sec:trex_galaxies}, the detection of a filamentary structure is more difficult: the \Trex algorithm detects a filament, but it only extends to $\sim \Rvir$, while in the same region the reconstructed map has its third-highest structure, but this does not meet the detection threshold. This might be again due to lower completeness in the area \citep[see][Fig. 9]{Owers2011-A2744_galaxies}.

Combining our results from X-ray and galaxy data, we can draw a comprehensive picture of the surroundings of the cluster A2744.  
From our analysis of A2744 on the plane of the sky, we identify three filaments connected to the cluster (NW, S, and E). These three are consistent across almost all the combinations of probe (X-rays or galaxies) and detection method (\Trex or azimuthal analysis), with larger uncertainty for the S filament in galaxy data, and also consistent with previous works. All three extend out to $\sim 1.5\, \Rvir$. In addition, analysing the 3D galaxy distribution with \Trex, we identified two filamentary structures along the line of sight, one in the front and one in the back of the cluster. The exact extent of these structures may however depend on the specific FoG correction we have applied to the galaxy sample. In total, we thus identify five filamentary structures connected to the core of A2744.

Such a large number of connected structures can help explain the particularly complex internal structure of the cluster. A2744 is known for its exceptional number of massive substructures, which make it a very dynamically disturbed cluster \citep{Merten2011-A2744_core_lensing, Owers2011-A2744_galaxies, Jauzac2015-A2744_structure, Jauzac2016-A2744_structure, Medezinski2016-A2744_structure, Bergamini2023-A2744_JWST_SL, Harvey&Massey2024-A2744_JWST_WL}. Cosmic filaments act as highways, along which gas and dark matter preferentially fall into clusters \citep{Rost2021-filaments_cluster-outskirts}, and thus can point to the origin and direction of some substructures. For example, \cite{Jauzac2016-A2744_structure} reported the alignment of three substructures in the direction of the NW filament. Another substructure in the north of the central cluster area shows evidence of a northward movement \citep{Owers2011-A2744_galaxies, Jauzac2016-A2744_structure}, suggesting a possible origin in the direction of the S filament. In a similar way, a shock front detected in the southeast of the cluster centre \citep{Owers2011-A2744_galaxies} suggests motion of substructures coming from the northwest.

Broadening the view, we compare our results with those on larger cluster populations. First of all, following \cite{Gouin2021} and \cite{Darragh-Ford2019}, we define the cluster connectivity $\kappa$ as the number of filaments connected to the cluster that cross a sphere of $1.5\, \Rvir$ radius. With this definition, taking into account the aforementioned uncertainties, we estimate the connectivity of A2744 to be in the range $\kappa \sim 3-5$.
It has been shown from both simulations and observations \citep{Aragon-Calvo2010, Codis2018, Darragh-Ford2019, Sarron2019, Kraljic2020, Lee2021, Gouin2021, Galarraga-Espinosa2023} that higher mass clusters tend to have higher connectivity. Moreover, \cite{Gouin2021} showed that, for the same mass, dynamically unrelaxed clusters are generally more connected than relaxed ones. 
A value of $\kappa\sim 5$ for the connectivity of A2744 would be well in agreement with trends from hydrodynamical simulations \citep{Gouin2021}.
Comparing to other observations of individual clusters, \cite{Einasto2020-A2142_filaments} found the dynamically disturbed cluster A2142 \citep[$M_{200}=1.2\times10^{15}\Msun$,][]{Munari2014} to be connected to $6-7$ filaments, higher than what we found for A2744, probably because A2142 lies inside of a supercluster, surrounded by many other groups. On the other hand, Coma \citep[$M_{200}=5.3\times10^{14}\Msun$,][]{Gavazzi2009} has lower connectivity, $\kappa = 3$ \citep{Malavasi2020, Malavasi2023}, which seems to be more in line with that of relaxed clusters from simulations \citep{Gouin2021}.

\section{Conclusions}\label{sec:conclusion}

We presented in this work the analysis of the X-ray and galaxy distributions surrounding the galaxy cluster A2744, and the identification of filamentary structures connected to the cluster. This is done using two methods: the aperture multipole decomposition, which highlights the azimuthal structure of the considered field in a given annulus; and the filament-finding method \Trex, which extracts the ridges in a discrete distribution using a statistical tree-based description.
We present here a brief summary of our results:

\begin{itemize}
    \item We reported for the first time the blind detection of filaments connected to a cluster from X-ray data: both methods we applied showed three clear filamentary structures in the S, NW and E of the cluster. These are consistent with previous visual detection by \cite{Eckert2015-A2744}.
    \item All three filaments are extracted from the distribution of galaxies at approximately the same (projected) positions as their X-ray counterparts, although the S filament is only clearly detected by the \Trex method.
    \item The X-ray-based \Trex probability map showed an additional structure in the north, coincident to one identified by \cite{Eckert2015-A2744}, but not confirmed by the multipole analysis nor by galaxy data.
    \item Through the galaxy distribution, we studied the 3-dimensional filamentary structure extracted with the \Trex method: we found that the three filaments detected in 2D are all almost perpendicular to the line of sight, but while the S and E filaments lie almost on the same plane, the NW one is found further in the back and connects to a different part of the cluster. The central part of the cluster is strongly elongated in the radial direction, and extends beyond $\Rvir$ both in the front and in the back. Furthermore, a loosely connected structure is identified in the back of the cluster, which seems to be aligned with the southern X-ray peak.
    \item The number and positions of the detected filaments can improve the interpretation of the highly disturbed cluster centre, in particular concerning the origin and direction of motion of its many substructures.
    \item We estimated the connectivity of A2744 to be in the range $\kappa\sim 3-5$, which is in agreement with trends from hydrodynamical simulations for a massive, disturbed cluster. 
\end{itemize}

With the combination of different methods for identifying filamentary structures, applied to different probes, we proved the possibility of blind detection of filaments in the outskirts of galaxy clusters. This result could open the way to a systematic search for cosmic filaments connected to clusters, in particular for what concerns the gas component, thanks to large X-ray programs such as eROSITA \citep{Bulbul2024-eROSITA_cluster_catalog}, CHEX-MATE \citep{CHEX-MATE2021}, and XRISM \citep{XRISM2020}.

\begin{acknowledgements}
The authors thank S. Ettori for discussions during the initial stage of the project. This work was supported by funding from the ByoPiC project from the European Research Council (ERC) under the European Union’s Horizon 2020 research and innovation program grant number ERC-2015-AdG 695561. SG acknowledges financial support from the Ecole Doctorale d’Astronomie et d’Astrophysique d’Ile-de-France (ED AAIF). CG acknowledges funding from the French Agence Nationale de la Recherche for the project WEAVEQSO-JPAS (ANR-22-CE31-0026).

\end{acknowledgements}

\bibliographystyle{aa}
\bibliography{references}

\appendix

\end{document}